\documentclass[12pt]{article}

\renewcommand{\title}[1]{\null\vspace{10mm}\noindent
                         {\Large{\bf #1}}\vspace{10mm}}
\newcommand{\authors}[1]{\noindent{\large #1}\vspace{5mm}}
\newcommand{\address}[1]{\center{\noindent\small\itshape #1\vspace{0mm}}}
\usepackage{bbm}

\begin{document}

\thispagestyle{empty}

\begin{titlepage}

\begin{center}
\hspace*{\fill}{{\normalsize \begin{tabular}{l}
                             \textsf{hep-th/0302038}\\
                             \textsf{TUW-03-05} 
                             \end{tabular}   }}

\title{Technical Remarks and
 Comments on the\\ UV/IR-Mixing Problem of a 
Noncommutative\\ Scalar Quantum Field Theory}

\authors {{F.~Aigner$^1$, M.~Hillbrand$^1$, J.~Knapp$^1$, 
G.~Milovanovic$^1$, V.~Putz$^2$, R.~Sch\"ofbeck$^1$, M.~Schweda$^3$}}  

\address{${}^{1,3}$Institut f\"ur Theoretische Physik, 
Technische Universit\"at 
Wien, \\ Wiedner Hauptstra\ss e 8-10, A-1040 Wien, Austria}
\address{${}^{2}$Max-Planck-Institute for Mathematics in the Sciences\\
Inselstra\ss e 22--26, D-04103 Leipzig, Germany}

\footnotetext[2]{vputz@mis.mpg.de, work supported by ``Fonds zur F\"orderung
der Wissenschaftlichen Forschung'' (FWF) under contract P15015-TPH.}

\footnotetext[3]{mschweda@tph.tuwien.ac.at}

\vspace{20mm}

\begin{minipage}{12cm}
  
  {\it Abstract.} In this letter we will dicuss the possibility of a 
resummation procedure in order to cure the UV/IR-mixing problem of
noncommutative field theories. The method is presented for a scalar
$\phi^4$ theory on Euclidean space. Finally, we sketch the idea of resummation
for $U(1)$-gauge theories.

\vspace*{1cm}
\end{minipage}

\end{center}

\end{titlepage}
              
\section{Introduction}

In the first part of the present paper we discuss different possibilities 
to cure the UV/IR-mixing of a scalar noncommutative quantum field theory at 
the one-loop level. Especially, we investigate the idea of resummation 
proposed in the literature \cite{Griguolo:2001ez}. In the second part we 
give some ideas how to import this method onto gauge theories.\\
In order to describe the noncommutative 
scalar quantum field theory one generalizes the usual concepts of
quantum mechanics, which are defined by the following commutation relation
\cite{Filk:dm}, \cite{Micu:2000xj}, 
\begin{equation}
 [\hat X_\mu,\hat P_\nu] = i\delta_{\mu\nu},\quad [\hat X_\mu,\hat X_\nu]
 = [\hat P_\nu, \hat P_\nu] = 0, \qquad\qquad (\hbar = 1)
\end{equation}
where $\hat X_\nu$ and $\hat P_\nu$ are the hermitian position and momentum
 operators, respectively. \\
However, there is no evidence that these concepts are sufficient at very 
short distance, implying the following natural generalization
\begin{equation}
 [\hat X_\mu,\hat X_\nu] = i\theta_{\mu\nu},
\end{equation}
with the deformation parameter $\theta_{\mu\nu}$ being an antisymmetric 
constant matrix of 
dimension [length]$^2$. In order to maintain Lorentz symmetry, the 
existence of such a constant antisymmetric tensor 
makes necessary a modification of the Lorentz transformation 
\cite{Bichl:2001yf}. To construct the perturbative 
field theory formulation we will use ordinary fields and not operator-valued
objects. One has the so-called Moyal-Weyl correspondence defined by
\begin{equation}   
\hat \phi(\hat X) \Longleftrightarrow \phi(x),
\end{equation}
where $\hat \phi(\hat X)$ is the operator valued functional and $\phi(x)$ 
is the usual scalar field depending on ordinary (commuting) Euclidean
coordinates $x_\mu$. The correspondence is given by
\begin{eqnarray}
\hat\phi(\hat X) &=& \int \frac{d^4k}{(2\pi)^4}e^{ik\hat X}\phi(k), \nonumber\\
\phi(k) &=& \int d^4x e^{-ikx}\phi{x}.
\end{eqnarray}
Here $k$ and $x$ are commutative, real variables. 
With the help of the Baker-Campbell-Hausdorff-formula one finds
\begin{eqnarray}
\hat\phi_1(\hat X)\hat\phi_2(\hat X) &=& \int\frac{d^4p}{(2\pi)^4}
\int\frac{d^4q}{(2\pi)^4} e^{ip\hat X}
e^{iq\hat X}\phi_1(p)\phi_2(q) \nonumber\\
&=& \int\frac{d^4p}{(2\pi)^4}\int\frac{d^4q}{(2\pi)^4}
e^{i(p+q)\hat X-\frac i2\theta_{\mu\nu}p_\mu q_\nu}\phi_1(p)\phi_2(q).
\end{eqnarray}
Now one can define the Moyal (or star)-product,
\begin{equation}
 \hat\phi_1(\hat X)\hat\phi_2(\hat X) \Longleftrightarrow \Big(\phi_1
 \star\phi_2\Big)(x),
\end{equation}
which is defined as \cite{Filk:dm}, \cite{Gracia-Bondia:1987kw}
\begin{equation}
\Big(\phi_1\star\phi_2\Big)(x) := \int \frac{d^4 k}{(2\pi)^4}\int d^4y 
\phi_1(x+\frac12\tilde k)\phi_2(x+y)e^{iky},\quad  \tilde k_\mu 
= \theta_{\mu\nu}k_\nu.
\end{equation}
With the star-product one is able to define the 
noncommutative scalar self-interacting classical action in a four-dimensional
Euclidean space,
\begin{equation}
 \label{action}
 \Gamma^{(0)}[\phi] = \int d^4x\Big(\frac12\big(\partial_\mu\phi(x)
 \partial_\mu\phi(x) +
 m^2\phi^2(x)\big) + \frac{g^2}{4!}\phi\star\phi\star\phi\star\phi(x)\Big).
\end{equation}
Note that the quadratic part of the action 
(\ref{action}) remains unchanged compared to the commutative theory.
Only the interaction term is modified and therefore the corresponding
Feynman-rules for the interaction vertices are changed by additional phase 
factors. In perturbation theory, the interplay of these phase factors 
produces two different types of graphs leading to planar and non-planar 
contributions. These are distinguished by their behaviour with respect
to the ultra-violet (UV)-region. The planar graphs show the desired 
effects expected from naive power counting and known from commutative theory.
The planar divergent radiative corrections can therefore be discussed in the
framework of the usual renormalization procedure. The second class of graphs,
the non-planar ones, show an ugly nonlocal behaviour. The a priori divergent
contributions are regularized by the phase factors. Thus, the non-planar
graphs are UV-finite, but instead develop a new singularity for 
small external momenta (if one discusses one-loop self-energy corrections).
This artefact is the so called UV/IR-mixing problem of noncommutative 
quantum field theories \cite{Minwalla:1999px}. The present work
rediscusses this UV/IR-problem within the framework of the resummed theory
\cite{Griguolo:2001ez} for a scalar $\phi^4$ theory and presents the
idea of a resummation of gauge field models.  

\section{Review of the traditional approach}

In order to understand the UV/IR-mixing in the framework of a resummed theory
it is useful to show how the UV/IR-problem enters the game 
\cite{Griguolo:2001ez}, \cite{Minwalla:1999px}. The action (\ref{action}) 
induces the following one-loop Feynman integral describing the first order 
quantum correction to the two point function
\begin{equation}
\Delta \Sigma = \frac{g^2}6\int \frac{d^4k}{(2\pi)^4} \frac1{k^2+m^2}
\big(2+\cos(k\tilde p)\big),\quad \tilde p_\mu = \theta_{\mu\nu}k_\nu,
\end{equation}
where we have used the non-resummed propagator for the scalar field,
\begin{equation}
 \Delta(k) = \frac1{k^2+m^2},
\end{equation}
and the corresponding Feynman rule for the noncommutative interaction vertex 
\cite{Minwalla:1999px}. This integral splits up in a planar contribution
(leading to the usual mass renormalization),
\begin{equation}
 \Delta\Sigma_p = \frac{g^2}3
 \int \frac{d^4k}{(2\pi)^4} \frac1{k^2+m^2},
\end{equation}
and a non-planar contribution,
\begin{equation}
 \Delta\Sigma_{np} = \frac{g^2}6
 \int \frac{d^4k}{(2\pi)^4} \frac1{k^2+m^2}
 e^{ik\tilde p}.
\end{equation}
With the usual techniques using Schwinger parametrization one gets for the 
non-planar expression after Gaussian integration
\begin{equation}
 \Delta\Sigma_{np} = \frac{g^2}6\frac{\pi^2}{(2\pi)^4}\int_0^\infty 
 \frac{d\alpha}{\alpha^2} e^{-\alpha m^2-\frac{\tilde p^2}{4\alpha}},
\end{equation}
where $\tilde p^2$ acts as regulator. With $\int_0^\infty 
\frac {d\alpha}{\alpha^2} \exp(-u\alpha - v/(4\alpha)) = 4\sqrt{(u/v)}
K_1(\sqrt{uv})$ for positive real part of $u$ and $v$ we find
\begin{equation}
\Delta\Sigma_{np} = \frac{g^2}6\frac{\pi^2}{(2\pi)^4} 
4\sqrt{\frac{m^2}{\tilde p^2}} K_1\left(\sqrt{m^2\tilde p^2}\right). 
\label{BesselK}
\end{equation}
With the expansion of the modified Bessel function $K_1(x) = \frac1x +
\frac x4(2\gamma -1 - 2\ln 2) + \frac x2 \ln x + {\cal O}(x^3)$ we find 
for small $\tilde p^2$
\begin{equation}
 \Delta \Sigma_{np}=\frac{g^2}{24\pi^2}\Big(\frac1{\tilde p^2} 
 +\frac{m^2}4
 \ln(m^2\tilde p^2) + \dots\Big), \label{divergence}
\end{equation}
where the dots stand for the terms remaining finite for $\tilde p^2 
\rightarrow 0$.
Thus, as suggested  in the introduction, in the non-planar section of 
noncommutative field theory the 
original UV-divergence of the commutative theory has been 
regularized by the momentum dependend cut-off $\tilde p^2$. In the commutative
limit $\theta_{\mu\nu}=\tilde p^2 = 0$ the divergence reappears. 
Unfortunately, even the regularized divergence causes troubles: 
The first term on the right hand side of (\ref{divergence})
gives rise to severe IR-singularities when inserted into higher order loop 
integrals $\int d^4 p$. This miraculous conservation of misery is 
called UV/IR-mixing of divergences.
After performing the ordinary mass renormalization (treating the planar 
correction) $M^2 = m^2 + \delta m^2$ the effective two point action to first 
order is
\begin{eqnarray}
\Gamma_2^{(1)} = \int \frac{d^4p}{(2\pi)^4}\phi(p)\phi(-p)
\Big(p^2+M^2 + \frac{g^2}{24\pi^2}\big(\frac1{\tilde p^2} +\frac{M^2}4
 \ln(M^2\tilde p^2) + \dots\big)\Big).
\end{eqnarray}
Now, in order to eliminate the UV/IR-mixing, one has to handle the 
horrible nonlocal $1/\tilde p^2$ term. In \cite{Grimstrup:2002nr} a field 
redefinition has been tried. Here we present another approach, a resummation 
procedure coming from quantum field theory at finite temperature and 
imported on noncommutative field theories by \cite{Griguolo:2001ez}.

\section{The resummation theory revisited}

The recipe to cure the UV/IR-mixing (at one-loop order) via resummation 
is given by adding
and subtracting to the classical action (\ref{action}) the term 
\cite{Griguolo:2001ez},
\begin{equation}
\frac{g^2}{24\pi^2}\int d^4x\frac12\phi(x)
\frac1{\tilde\partial^2}\phi(x),
\end{equation}
implying that one has now the following tree level action,
\begin{equation}
\Gamma^{(0)}_R[\phi] = \int d^4x\Big(\frac12\big(\partial_\mu\phi(x) 
\partial_\mu\phi(x) +
m^2\phi^2(x) -\phi(x)\frac{\tilde c}{\tilde\partial^2}\phi(x) + 
\phi(x)\frac{\tilde c}{\tilde\partial^2}\phi(x)\big)
+ \frac{g^2}{4!}\phi\star\phi\star\phi\star\phi(x)\Big), \label{resact}
\end{equation}
with $\tilde c = \frac{g^2}{24\pi^2}$. The idea is to treat one of the
two cancelling terms as modification of the propagator and the other
as a two-point vertex function. Doing loop-expansion, this leads to a 
mixing of orders in the coupling constant $g^2$ (which is exactly the desired 
effect of the resummation procedure), but non-perturbatively the theory 
remains the same. The process of resummation allows in principle two 
possibilities for the resummed propagators,
\begin{equation}
\Delta_{\pm}(k)=\frac1{k^2+m^2\pm\frac{\tilde c}{\tilde k^2}}.\label{prop}
\end{equation}
As argued in \cite{Griguolo:2001ez} the negative sign corresponds to 
unphysical tachyonic poles. Therefore it seems natural that only the 
positive sign is meaningful. However, aiming at further calculations
concerning gauge theory, we want to stay as general as possible.\\
Now one can compute the one-loop quantum correction with the resummed 
propagator,
\begin{equation}
 \Delta\Sigma_\pm = \frac{g^2}6\int 
\frac{d^4k}{(2\pi)^4} \frac1{k^2+m^2
 \pm\frac{\tilde c}{\tilde k^2}}(2+\cos(k\tilde p)) 
 =\frac{g^2}6\int \frac{d^4k}{(2\pi)^4} \frac{\tilde k^2}
 {k^2\tilde k^2 +m^2\tilde k^2 \pm \tilde c}(2+e^{ik\tilde p}).
\label{resloop}
\end{equation}
In the UV-region the integral (\ref{resloop}) has the same structure as for 
the non-resummed theory. Thus, we expect a quite similar result, 
with a quadratically divergent contribution from 
the planar graph and a finite but nonlocal $\frac1{\tilde p^2}$-contribution
from the non-planar graph. This will be verified by explicit calculation.\\
In order to get a Gaussian integral after Schwinger parametrization we 
have to expand  (\ref{resloop}) into partial fractions. 
Unfortunately, the term 
$k^2\tilde k^2$ causes troubles unless $\tilde k^2 \propto k^2$. Since
$\theta_{\mu\nu}$ can always be transformed into a block matrix 
\begin{eqnarray}
 \theta_{\mu\nu} = \left(\begin{array}{cccc}
 0 &\theta_{12} & 0 & 0 \\
 -\theta_{12} &  0 & 0& 0\\
 0 & 0 & 0 & \theta_{34} \\
 0 & 0 & -\theta_{34} & 0
 \end{array}\right),
\end{eqnarray}
we see that $\theta_{\mu\nu}$ has only two degrees of freedom. By eliminating
one degree of freedom, thus using the choice $\theta_{12} = 
\theta_{34} =:\theta$, we find $(\theta^2)_{\mu\nu} = \theta^2\otimes 
\mathbbm{1}_{4\times4}$.
At least for this special choice we can calculate
\begin{eqnarray}
 \Delta\Sigma_{\pm}&=&\frac{g^2}6\int \frac{d^4k}{(2\pi)^4}\frac{
 k^2}{(k^2)^2+m^2k^2 \pm \frac{\tilde c}{\theta^2}}
 (2+e^{ik\tilde p})\nonumber\\
 &=& -\frac{g^2}6\int \frac{d^4k}{(2\pi)^4}\frac{k^2}{2u}
 \Big(\frac1{k^2 +\frac{m^2}2 + u} -  \frac1{k^2 +\frac{m^2}2 - u}\Big)
 (2+e^{ik\tilde p}).
\end{eqnarray}
Here the (possibly complex) quantity $u = \sqrt{\frac{m^4}{4}\mp\frac
{\tilde c}{\theta^2}}$. Now we can use formulae
(\ref{int2}), (\ref{int3}) from the 
appendix (if $u$ is real, we introduce a convergence factor $b\rightarrow
0$ by hand) and obtain for the non-planar part
\begin{eqnarray}
\Delta\Sigma_{np\pm}&=&\frac{g^2}{24\pi^2}
\frac1{2u}\Big((\frac{m^2}2 + u)
\sqrt{\frac{\frac{m^2}2 + u}{\tilde p^2}}K_1\left(\sqrt{(\frac{m^2}2 + u)
\tilde p^2}\right) \nonumber\\&&\quad
-(\frac{m^2}2 - u)
\sqrt{\frac{\frac{m^2}2 - u}{\tilde p^2}}K_1\left(\sqrt{(\frac{m^2}2 - u)
\tilde p^2}\right)\Big). \label{result1}
\end{eqnarray}
For $\tilde c = 0 \Rightarrow u = \frac{m^2}2$ this yields exactly the 
result (\ref{divergence}). 
With the expansion of the modified Bessel function we find for the 
$(\tilde p^2 \rightarrow 0)$-divergent part
\begin{eqnarray} 
\Delta\Sigma_{np\pm}&=&\frac{g^2}{24\pi^2}
\Big(\frac1{\tilde p^2} + \frac1{8u}(\frac{m^2}2+u)^2\ln((\frac{m^2}2+u)
\tilde p^2) \nonumber\\&&\qquad
- \frac1{8u}(\frac{m^2}2-u)^2\ln((\frac{m^2}2-u) \tilde p^2)  + \dots\Big),
\end{eqnarray}
where the dots denote the terms finite for $\tilde p^2\rightarrow 0$.
The logarithmic term is harmless, but we have to keep an eye on the 
$\frac1{\tilde p^2}$ term. We find that this term is invariant with respect 
to the choice of sign in (\ref{prop}). Therefore it is cancelled by the 
counterterm in the action only if we choose the positive sign in 
(\ref{prop}), so the counterterm reads
\begin{equation}
\delta\Gamma =  + \int d^4x \phi(x)\frac{\tilde c}{\tilde\partial^2}\phi(x)
\end{equation}
(note that $\tilde \partial^2 \Rightarrow -\tilde k^2)$.
The result of the planar part is obtained by 
multiplying (\ref{result1}) by 2 and taking the limes
$\tilde p \rightarrow 0$. 
\begin{eqnarray} 
\Delta\Sigma_{p\pm}&=&\frac{g^2}{12\pi^2}
\lim_{\Lambda\rightarrow \infty}
\Big(\Lambda^2 - \frac1{8u}(\frac{m^2}2+u)^2\ln\left(\frac{\Lambda^2}
{\frac{m^2}2+u}\right) \nonumber\\&&\qquad
+ \frac1{8u}(\frac{m^2}2-u)^2\ln\left(\frac{\Lambda^2}
{\frac{m^2}2-u} \right)  + \dots\Big),
\end{eqnarray}
where again the finite terms are ignored. Thus, also in resummed
field theory the planar one-loop correction of the two-point function
can be absorbed in an ordinary mass renormalization of the theory.

\section{Resummation in $U(1)$-noncommutative YM-theory?}

In this section we try to sketch the idea of a resummation procedure
for a gauge field model with BRS-symmetry
\cite{Becchi:1975nq}. We begin with a pure $U(1)$-noncommutative
Yang-Mills (NCYM)-theory, which is described in Euclidean space
at the classical level by
\begin{equation}
\Gamma^{(0)}_{INV} = \frac14\int d^4x F_{\mu\nu}\star F_{\mu\nu},
\end{equation}
where the field strength $F_{\mu\nu}$ is
\begin{equation}
F_{\mu\nu} = \partial_\mu A_\nu -\partial_\nu A_\mu -ig[A_\mu,A_\nu]_M, 
\quad [A,B]_M := A\star B - B\star A .
\end{equation}
The enclosure of fermions can be done in the usual way \cite{Hayakawa:1999zf},
\cite{Matusis:2000jf}. Scalar matter fields are treated in 
\cite{VanRaamsdonk:2001jd}. In order to allow a meaningful perturbation theory
one has to use the BRS-quantization procedure \cite{Becchi:1975nq}, which
implies the introduction of ghost fields $c,\bar c$ and a multiplier field
$B$ for the gauge fixing. One has the total action
\begin{eqnarray}
\Gamma^{(0)} &=& \Gamma^{(0)}_{INV} +\Gamma_{gf}+\Gamma_{matter} \nonumber\\
&=&\frac14\int d^4x F_{\mu\nu}\star F_{\mu\nu} + \int d^4x(gB\star\partial_\mu
A_\mu +\frac\alpha2 B\star B- 
\bar c\star \partial_\mu D_\mu c) +\Gamma_{matter},
\end{eqnarray}
where $D_\mu := \partial_\mu - ig[A_\mu,.]_M$. The corresponding 
BRS-transformation is given by
\begin{eqnarray}
 sA_\mu = D_\mu c,&\quad& sc = ic\star c, \nonumber\\
 s\bar c = B, &\quad& sB = 0. \label{BRS}
\end{eqnarray}
Doing now perturbation theory at the one-loop level (including all 
contributions coming from the gluon, ghost, fermion etc. fields) one obtains
the following problematic contribution to the vacuum polarization of the 
photon \cite{Hayakawa:1999zf},
\begin{equation}
\Pi_{\mu\nu}(k) = \beta\frac{\tilde k_\mu
\tilde k_\nu}{(\tilde k^2)^2} 
\end{equation}
(with $\beta$ some numerical constant of order $g^2$), which represents
the non-planar 
one-loop contribution. This is the well-known UV/IR-mixing term for 
non-supersymmetric NCYM-models. It is singular for $k_\mu\rightarrow 0$.
The corresponding term for resummation would be
\begin{equation}
\delta\Gamma=\beta\int \frac{d^4k}{(2\pi)^4}A_\mu(k)A_\nu(-k)\frac{\tilde k_\mu
\tilde k_\nu}{(\tilde k^2)^2}. \label{gaugecounter}
\end{equation}
As observed in \cite{VanRaamsdonk:2001jd} it is gauge-invariant with respect
to an infinitesimal Abelian gauge transformation, $\delta A_\mu =
\partial_\mu\lambda$. But this is not the full story, one has to respect
also BRS-invariance (\ref{BRS}). In order to transform (\ref{gaugecounter})
in a BRS-invariant quantity we first define
\begin{equation}
\tilde F := \theta_{\mu\nu}F_{\mu\nu},\quad \tilde D_\mu := 
\theta_{\mu\nu}D_\nu.
\end{equation}
The idea is to replace (\ref{gaugecounter}) by (written in Euclidean coordinate
space)
\begin{equation}
\delta\Gamma_{INV}=+\frac\beta4\int d^4 x
\tilde F\star\frac1{(\tilde D^2)^2}\star\tilde F.
\label{newcounter}
\end{equation}
Of course, $\frac1{(\tilde D^2)^2}$ is meant as power series in the gauge field
$A_\mu$, so we obtain an infinite set of nonlocal vertices (however, to each
order in the gauge coupling $g$ only a finite number of these vertices
contribute).
To lowest order in $A_\mu$ (\ref{gaugecounter}) and (\ref{newcounter}) are
identical. Indeed, it is easy to show that $\frac1{(\tilde D^2)^2}\star X$
transforms covariantly if $X$ does. For $X = \tilde F$ one has
\begin{equation}
\delta_\lambda \tilde F = i[\lambda, \tilde F]_M \Longrightarrow \quad
\delta_\lambda(\frac1{(\tilde D^2)^2}\star\tilde F) =
i[\lambda,\frac1{(\tilde D^2)^2}\star\tilde F]_M, 
\end{equation}
implying that (\ref{newcounter}) is BRS-invariant. In order to get a resummed
gauge field model one generalizes now the calculation of the last section. The
resummed action reads
\begin{eqnarray}
\Gamma^{(0)}_R &=& \int d^4x\Big(\frac14 F_{\mu\nu}\star F_{\mu\nu} + 
gB\star\partial_\mu A_\nu +\frac\alpha2 B\star B - 
\bar c\star\partial_\mu D_\mu c 
\nonumber\\ &&
\qquad+\frac\beta4 \tilde F\star\frac1{(\tilde D^2)^2}\star\tilde F 
-\frac\beta4\tilde F\star\frac1{(\tilde D^2)^2}\star\tilde F \Big)
+\Gamma_{matter}
\label{resggact}
\end{eqnarray}
A similar ansatz for the solution of an analogous problem
in high temperature QCD can be found in \cite{Kreuzer:ge}. Taking only the
bilinear part of (\ref{resggact}) one can calculate the resummed U(1)-gauge
field propagator as 
\begin{equation}  
\Delta_{\mu\nu\pm}(k) =-\frac1{k^2}\Big(g_{\mu\nu}-(1-\alpha)
\frac{k_\mu k_\nu}{k^2}
\mp2\beta\frac{\tilde k_\mu\tilde k_\nu}{(\tilde k^2)^2(k^2\pm\frac{2\beta}
{\tilde k^2})}\Big).
\end{equation}
The upper sign corresponds to the inclusion of the positive 
$\beta$-term in (\ref{resggact}) in the propagator 
(treating the negative $\beta$-term as counterterm).
One observes that the new term in the resummed propagator is independent of 
the gauge parameter $\alpha$. Moreover, it is transversal
due to $k_\mu\tilde k_\mu = 0$. The correct sign must be checked by explicit
 one-loop calculations \cite{done}. Note that the resummation procedure for
gauge theories involves also a resummation of the vertices due to the
non-bilinear parts in the resummed term in (\ref{resggact}). 

\section{Conclusion}

In this short note we have reinvestigated the resummation procedure for 
scalar noncommutative quantum field theory. We have discussed the two
possibilities of sign combination in the resummed action (\ref{resact}). It 
was verified by one-loop calculations that only
\begin{equation}
\Delta_+(k) = \frac1{k^2 +m^2+\frac{\tilde c}{\tilde k^2}}
\end{equation} 
is useful in order to compensate the IR-singularity of $\frac1{\tilde k^2}$.
Encouraged by works in thermal quantum field theory \cite{Kreuzer:ge},
\cite{Fischler:2000fv} we have proposed a generalization for gauge field 
models. However, one has to stress that the relevant calculations have still
to be done \cite{done}.

\section{Acknowledgements}

We would like to thank M. Wickenhauser, R. Wulkenhaar and H. Grosse for 
helpful discussion concerning the tremendous difficulties with integrals.

\appendix

\section{Integrals}

First we need the complex Euclidean Gauss integral ($\alpha$ real)
\begin{eqnarray}
&&\int d^4k e^{\pm i \alpha k^2 +ik\tilde p} = 
\int d^4k e^{\pm i\alpha(k\pm\frac{\tilde p}{2\alpha})^2
\mp\frac{i\tilde p^2}{4\alpha}}
=\nonumber\\&&\quad=
\lim_{\varepsilon\rightarrow 0}\int d^4k' e^{-(\varepsilon\mp i\alpha)k'^2 \mp
\frac{i\tilde p^2}{4\alpha}}
=-\frac{\pi^2}{\alpha^2}
e^{\mp\frac{i\tilde p^2}{4\alpha}}.\label{int1}
\end{eqnarray}
With this we get for real $b>0,\ a$ real,
\begin{eqnarray}
&&\int d^4k \frac{k^2}{k^2 +a \pm ib}e^{ik\tilde p}=\nonumber\\&&\quad
=\int d^4k(\pm i)\int_0^{\infty}
d\alpha (\pm i\frac\partial{\partial\alpha}+a \pm ib)e^{\pm i\alpha(k^2 + a
\pm ib) + ik\tilde p} \nonumber\\&&\quad
=\pm i \int_0^{\infty}d\alpha (\pm i\frac\partial{\partial\alpha}+a \pm ib)
\big(-\frac{\pi^2}{\alpha^2}e^{\mp\frac{i\tilde p^2}{4\alpha}\pm i\alpha(a\pm 
ib)}\big) \nonumber\\&&\quad
 = -(\pm i)\pi^2(a \pm ib)\lim_{\varepsilon\rightarrow 0}
\int_0^\infty \frac{d\alpha}{\alpha^2}
e^{-\frac{\varepsilon\pm i\tilde p^2}{4\alpha}-(b\mp ia)\alpha}.\label{int2}
\end{eqnarray}
Here we use $\int_0^\infty 
\frac {d\alpha}{\alpha^2} \exp(-u\alpha - v/(4\alpha)) = 4(\sqrt{u}/
\sqrt{v}) K_1(\sqrt{uv})$ for positive real part of $u$ and $v$ and find
\begin{eqnarray}
&&
=-(\sqrt{\pm i})^24\pi^2(a\pm ib)\frac{\sqrt{b\mp ia}}{\sqrt{\pm i\tilde p^2}}
K_1\big(\sqrt{\pm i\tilde p^2(b\mp ia)}\big) \nonumber\\ &&
=-4\pi^2(a\pm ib)\sqrt{\frac{a\pm ib}
{\tilde p^2}}K_1\big(\sqrt{(a\pm ib)\tilde p^2}\big). \label{int3}
\end{eqnarray}
Of course one has to be careful with respect to the sign of the roots, thus 
one should check the result (\ref{int3}) for special cases (e.~g. $a >0,\ 
b\rightarrow 0$) via ordinary (non-complex) Schwinger parametrization.

\end{document}